%% file: phishcatch.tex
\RequirePackage{amsmath}
\documentclass[runningheads]{llncs}

\usepackage{hyperref}
\usepackage[utf8]{inputenc}
\usepackage[T1]{fontenc}
\usepackage[english]{babel}
\usepackage[babel=true]{microtype}

\usepackage{graphicx}
\usepackage{color}
\usepackage[dvipsnames]{xcolor}

\usepackage{mathrsfs}
\usepackage{amssymb}
\usepackage{bm}
\usepackage{dsfont}

\input{preamble-commands}

\begin{document}

\title{%
  Catching the Phish: Detecting Phishing Attacks using Recurrent Neural Networks (RNNs)%
  %\texorpdfstring{\thanks{Supported by organization x.}}{}%
}
\titlerunning{Catching the Phish}

\author{%
  Luk\'{a}\v{s} Halga\v{s} \inst{1} \and
  Ioannis Agrafiotis \inst{1} \and
  Jason R.C.~Nurse \inst{2}%
}

\authorrunning{L.~Halga\v{s}~\etal}

\institute{%
  Department of Computer Science,
  University of Oxford,
  United Kingdom
  % \\
  % Wolfson Building, Parks Road OX1 3QD, Oxford, UK
  \email{\texorpdfstring{%
    \{lukas.halgas,%
      ioannis.agrafiotis%
    \}%
    @cs.ox.ac.uk%
  }{}} \and
  School of Computing,
  University of Kent,
  Canterbury,
  United Kingdom
  \email{\texorpdfstring{%
    j.r.c.nurse@kent.ac.uk%
  }{}}%
}

% typeset the header of the contribution
\maketitle

\input{0-abstract}
\input{1-introduction}
\input{2-background}
\input{3-methodology}
\input{4-design-implementation}
\input{5-evaluation.tex}
\input{6-conclusion.tex}

\bibliographystyle{splncs04}
\bibliography{phishcatch}

\end{document}

%% file: preamble-commands.tex
\newcommand{\eg}{e.g.}
\newcommand{\ie}{i.e.}
\newcommand{\etal}{et~al.}

\DeclareMathOperator*{\argmax}{arg\,max}
\newcommand{\1}{\mathds{1}}
\newcommand{\xraw}{\bm{x}_{\mathrm{raw}}}

\newcommand{\jnfull}{\textsf{SA-\!JN}}
\newcommand{\enjn}{\textsf{En-\!JN}}
\newcommand{\pnjn}{\textsf{?-\!JN}}

\newcommand{\dmctext}{\ensuremath{\mathrm{DMC}_{\mathrm{text}}}}
\newcommand{\nettext}{\ensuremath{\mathrm{textAnalysis}}}

%% file: 0-abstract.tex
\begin{abstract}

The emergence of online services in our daily lives has been accompanied by a
range of malicious attempts to trick individuals into performing undesired
actions, often to the benefit of the adversary. The most popular medium of
these attempts is phishing attacks, particularly through emails and websites.
In order to defend against such attacks, there is an urgent need for automated
mechanisms to identify this malevolent content before it reaches users. Machine
learning techniques have gradually become the standard for such classification
problems. However, identifying common measurable features of phishing content
(e.g., in emails) is notoriously difficult. To address this problem, we engage
in a novel study into a phishing content classifier based on a recurrent neural
network (RNN), which identifies such features without human input. At this
stage, we scope our research to emails, but our approach can be extended to
apply to websites. Our results show that the proposed system outperforms
state-of-the-art tools. Furthermore, our classifier is efficient and takes into
account only the text and, in particular, the textual structure of the email.
Since these features are rarely considered in email classification, we argue
that our classifier can complement existing classifiers with high information
gain.

\keywords{%
  Phishing \and
  Machine Learning \and
  Recurrent Neural Networks \and
  Natural Language Processing \and
  Web security.%
}
\end{abstract}

%% file: 1-introduction.tex
\section{Introduction}

Advances in computer security have raised confidence in internet safety leading
to e-commerce, internet banking and other means of sending, managing and
receiving money online. Unfortunately, the advent of online services has been
accompanied by illicit attempts to sham such transactions to the benefit of
malicious entities. Perhaps the most popular and easy to execute attack, which
poses a threat to organisations, institutions and simple users, is
\emph{phishing}.

Phishing is a type of cyber-attack that communicates socially engineered
messages to humans using digital channels in order to persuade them to perform
certain activities to the attacker's
benefit~\cite{khonji2013survey,nurse2019cybercrime}. Email is the most popular
avenue for a phishing attack, with almost 91\% of successful
cyber-attacks/security breaches initiated by sending out spoofed
emails~\cite{phishme2016report}. The entire phishing operation can even be
outsourced and automated~\cite{ramzan2010countermeasures}, enabling the
phishing threat to be, as it is, ubiquitous and continuous. Research has also
found that it is increasingly difficult for humans to detect phishing
attacks~\cite{iuga2016hook}. Therefore, there is a strong argument for
automated mitigation methods to keep the user's exposure to the attacks at a
minimum.

The dynamic nature of phishing, with new trends and challenges constantly
emerging, motivates a more adaptive filtering approach. Machine learning (ML)
has been utilised, as the de-facto standard for classification purposes over
many fields, email classification included. Developing a ML-based classifier to
underline the phishing filtering, is the approach we investigate in this paper.
Our classifier analyses the text of the email and, in particular, the email's
language structure. It follows that our work is largely orthogonal to
contemporary email classification systems which, to the best of the authors'
knowledge, do not employ natural language processing. We propose a novel
detection system for phishing emails based on recurrent neural networks (RNNs).
Our evaluation indicates that the RNN system outperforms state-of-the-art
tools.

In what follows, Section~\ref{sec:background} presents alternative approaches
to automated detection of phishing emails and literature on current machine
learning approaches. Section~\ref{sec:methodology} details our methodology and
feature selection for the RNN, while Section~\ref{sec:design} describes the
system implementation. Section~\ref{sec:evaluation} discusses the evaluation of
the system and Section~\ref{sec:conclusion} concludes the paper.

%% file: 2-background.tex
\section{Current landscape on mitigation techniques to phishing}
\label{sec:background}

Specialised algorithms to classify email as phishing, spam (unsolicited email)
or ham (\ie, not spam) have been the focus of research since the beginning of
unsolicited email. Filtering of phishing emails is subsumed in the more general
problem of spam filtering. As such, most email classifiers, hence filters,
treat relatively harmless spam equivalently to dangerous phishing emails.

Chandrasekaran et al., propose that phishing filtering needs to be treated
separately from the bulk spam filtering~\cite{chandrasekaran2006structural}.
Phishing emails mimic ham emails, in that they want to raise confidence in its
perfectly legitimate origin. Their proposed classifier uses 23 \emph{style
marker features} including the number of words $W$, number of characters $C$
and number of words per character $W/C$ aliased as \emph{vocabulary richness}.
The authors report results of up to perfect classification, with the accuracy
dropping by 20\% with the removal of the two structure features. Although the
experiment used only a small corpus of 400 emails, the results demonstrate the
importance of language, layout and structure of emails in phishing
classification.

PILFER is an ML-based phishing email classifier proposed by
Fette~\etal~\cite{fette2007learning}. The authors identified a subset of only
ten (from the hundreds of features popularly used to classify spam) that best
distinguish phishing emails from ham. PILFER outperformed the trained version
of Apache's SpamAssassin~\cite{spamassassin} at classifying phishing emails,
with the false negative rate reduced by a factor of ten. This result
demonstrates that having specialised features for the task in prominence to
general email classification features improves phishing classification.

Bergholz~\etal~\cite{bergholz2008model-based} build on top of this work by
introducing advanced features for phishing classification. The authors note the
statistical insignificance of classification improvement through variation of
the classifying algorithm itself. They conclude that a statistically
significant improvement is possible by invention of better features. Bergholz
et al. develop two sets of advanced features based on unsupervised learning
algorithms to complement 27 basic features commonly used in spam detection. The
advanced features are the Dynamic Markov Chain (DMC) model features, and Latent
Topic Model Features: word-clusters of topics based on Latent Dirichlet
Allocation (LDA). The best results occur when the advanced features are used in
conjunction with the basic features, achieving the
state-of-the-art~\cite{bergholz2008model-based}.

Toolan~\etal~\cite{toolan2010feature-selection} analysed 40 basic features
popularly used in email classification and ranked them based on their
\emph{information gain} to the classification task at hand. The most
informative features were vocabulary richness of the email body and of the
subject. Other popular features performed very poorly, indicating that our
intuitive understanding of what constitutes a phishing email may be very wrong.
This is illustrated by the failure to gain information from counting
\textsf{<form>} elements or finding the word `debit' in the subject. We may
attribute the results to a shift in phishing trends, or to the failure of human
experts to identify good features. The authors also conclude that language
modelling approaches to phishing classification are the most promising.

%% file: 3-methodology.tex
\section{Methodology}
\label{sec:methodology}

The \emph{classification} task is to identify to which of a finite number $k$
of categories, or classes $C = \{c_1, \ldots, c_k\}$, a sample $\bm x$ belongs,
\ie, deduce a classifier or mapping, $\bm x \mapsto c$. In our application to
phishing, the classification is a representation of an email to the label set
$\{\mathit{phish}, \mathit{ham}\}$.

In the task of email classification for the exclusion of emails from delivery,
we emphasise on \emph{precision} which is defined in
Section~\ref{sec:evaluation}, as a criterion of a successful classification.
Our rational being that it is essential to elevate the importance of correctly
classifying ham emails above classifying phishing emails as phish. This
challenge of false positives, or misclassifying ham email as phishing, is the
main reason for users' resistance to email filters.

The machine learning approach to classification is to automatically establish a
function $f$ that determines the desired class $$ \widehat{y} = f(\bm x) \in
\{\mathit{ham}, \mathit{phish}\} $$ on the input of a representation $\bm x$ of
an email. The function $f$ is parameterized by values $\bm \theta$. During the
training phase, the parameter values $\bm \theta$ are determined to reproduce a
relation between the input $\bm x$ and class label $y$ in agreement with a
training set $\{ (\bm x_0, y_0), \ldots, (\bm x_n, y_n) \}$ of pre-classified
samples and a suitable optimisation criterion. In this sense, the ML approach
is to extrapolate this relation between the observed sample points and class
labels to unlabelled input $\bm x$ and its predicted class $\hat{y} = f(\bm
x)$. To enforce the precision requirement, we encode excess penalty for false
positives in the optimisation criterion, skewing the potential
precision/accuracy trade-off towards the classification of ham emails.

\subsection{Feature Identification}

An input $\xraw$ representing an email as a (very long) series of binary
digits, comprising the raw source code of an email in binary format, is
unwieldy for an algorithm to detect patterns. We hence use a more compact
representation of the input as a feature vector $\bm{x} =
\big(f_1(\xraw),\ldots,f_m(\xraw)\big)$. Features should characterise an email
with respect to the current classification problem. The relative inaccuracy of
ML-based spam classifiers on the seemingly similar task of phishing
classification illustrates the need for specialised features for this task.

Features are most often identified by experts, in line with their intuitive
understanding of ``phishiness'' or ``hamness''.
Toolan~\etal~\cite{toolan2010feature-selection} demonstrated that such
intuitively sound features often fail to inform the classification under
discussion. On the other hand, structural features have empirically been
indicative of emails being ham or phish~\cite{fette2007learning}. Based on
this, we therefore follow the language modelling approaches to the challenge of
phishing classification which are viewed as the most
worthwhile~\cite{toolan2010feature-selection}.

Natural Language Processing (NLP) is the field of Computer Science studying
human-machine interactions and, in particular, establishing and exploiting
language models. The rich structure and ambiguity of natural languages make it
difficult to identify and extract complex language features, such as the tone
of urgency in the email body. As explained in Section~\ref{sec:background},
Verma~\etal~\cite{verma2012detecting} used pre-trained WordNet hypernymy trees
of sets of words conveying urgency or action, among other characteristics, to
identify sentences and hence emails as actionable or informative.

In the unsupervised learning approach, the ML algorithm detects data patterns
in the dataset without supervision or explicit expert advice. That is, the
training of the model determines, or learns, the features itself.
Bergholz~\etal~\cite{bergholz2010new} trained a dynamic Markov chain language model
to generate ham or phishing emails. We utilise similar NLP techniques in our
system.

\subsection{Deep Learning}

Neural Networks (NNs) are a computational model, in the quintessential example
of a multilayer perceptron (MLP) resembling a hierarchical network of units, or
neurones. The hierarchical structure intuitively gives NNs the capacity to
extract high-level features from simple data, \ie, to disentangle and winnow
the factor of variation in the NN input. This intuition of NN structure makes
NNs suitable for the task of \emph{representation learning}, or automatic
feature identification.

Recurrent Neural Networks (RNNs), the deepest of all learners, are a family of
NNs specialised for processing sequential data. Like Markov chain models, RNNs
have the advantage of processing data in sequence, thus accounting for the
order of data. The input text is usually abstracted to a sequence of
characters, words or phrases. Undoubtedly, the order of words is valuable in
language modelling. RNNs form the backbone of current state-of-the-art language
models, so an RNN language model could form an accurate content-based
classifier of emails.

It is worth mentioning that RNNs have been applied by previous works to address
the problem of classifying malicious URLs and websites. Researchers have used
various features and subsequently classified with high accuracy websites and
URLs into malware, phishing and
benign~\cite{bahnsen2017url-rnn,mohammad2014predicting,vinayakumar2018urls,%
zhao2018classifying}. We extend the classification problem by considering only
language models for phishing emails.

We alleviate the learning problem from language modelling to the binary
classification of email to phish or ham. This classification can be trivially
abstracted to predicting $y$, where $y = 1$ if \textit{phish} and $y = 0$ if
\textit{ham}. We thus get a supervised learning problem with representation
learning. This simpler task overcomes the often-prohibitive computational cost
of training a full-blown language model. Inherently, the RNN classifier models
a $y \sim \mathrm{Bernoulli}(p_{\bm x})$ distribution using a sigmoid output
unit
\begin{equation*}
  p_{\bm x} = \sigma(z_{\bm x})
           := \frac{1}{1 + \exp(-z_{\bm x})}
            = \frac{\exp(z_{\bm x})}{\sum_{y'=0}^{1} \exp(y'\,z_{\bm x})}
\end{equation*}
where $z_{\bm x}$ is the output of the last linear layer, dependent on the RNN
input $\bm x$. Intuitively, this is the normalisation of the unnormalised
probability distribution
\begin{align*}
       \widetilde p_{\bm x}(c) &= \exp(c\,z_{\bm x}) \\
  \log \widetilde p_{\bm x}(c) &= c\,z_{\bm x}
\end{align*}
for $c \in \{0,1\}$. Then
$
  p_{\bm x}
  = p(y = 1 \mid {\mbox{\scriptsize{sequence of words of email}}\ \bm x)}
  \in [0,1]
$
gives the email label prediction
$
  \widehat y
  = \argmax_{c \in \{0,1\}} p(y = c \mid \bm x)
  = \1\{z_{\bm x} \geq 0\}
$.

%% file: 4-design-implementation.tex
\section{Design and Implementation}
\label{sec:design}

Our RNN classifier labels an input email as either a legitimate email or a
phishing attempt. In this section, we describe the procedure of transforming
the raw email source into a variable size vector of integers that is input to
the RNN itself.

\subsection{Preprocessing for the RNN Classifier}

Our binary classification RNN model takes sequences of integer values as input
and outputs a value between 0 and 1. We abstract the computer-native copy of an
email as a sequence of bytes into the high-level representation as a sequence
of symbol and word tokens, represented as unique integers. It is customary to
`feed' RNNs with an $n$-gram representation of the abstracted text. Due to the
small size of our dataset, our dictionary of $n$-grams would contain very few
repetitive phrases of $n$ words for values $n \geq 2$. For the balance of token
expressiveness, and vocabulary size, we choose to represent emails as sequences
of 1-grams, or single-word tokens.

Note that our classifier only considers the text of emails in making its
classification decision. Thus, effective features, such as those based on
linked web address analysis, are completely orthogonal to our classifier and
thus are largely complementary. As an initial step in preprocessing of the
classified email, we extract its text in plaintext format.

\subsection{Tokenizing the Text}

We seek flexibility in tokenizing the text through fine tuning the parameters
of the tokenizer, such as rules of what word or character sequences to
represent as the same token. The na\"ive approach of splitting on whitespace
characters does not generalise well to email tokenizing. Incautious or
malicious salting, \eg, inconsistent whitespace or the ubiquity of special
characters, form words unique to an email. Considering such tokens would
inherently lead to overfitting, based on the presence of unique traits.

Our approach to tokenizing is that of adjusted word-splitting. First, we
lowercase all characters in the email and remove all characters the RFC 3986
standard does not allow to be present in a URL, \ie, we only keep the
unreserved \texttt{a-z}, \texttt{0-9},
\texttt{-\,.\,\_\,\raisebox{-0.5ex}{\textasciitilde}} and reserved
\texttt{:\,/\,?\,\#\,[\,]\,@}\,\texttt{!\,\$\,\&\,'\,(\,)\,*\,+\,,\,;\,=}\
characters and the percentage sign \texttt{\%}. Although this step is motivated
by ease of later identifying URLs for the \texttt{<url>} token determination,
we get the benefit of restricting our character base cardinality to 61. The
60th character, which RFC 3986 does not allow in URL but we do not immediately
replace with whitespace, is the quote character \texttt{"}, which is often used
in emails. Note, the 61st character is the whitespace character.

We introduce four special tokens summed up in Table~\ref{tbl:tokens}, and, nine
tokens for the special characters left, replacing dots, quotes and seven other
special characters with their respective tokens. Finally, we split the clean
text into words, serving as their individual tokens, and prepend and append the
start \texttt{<s>} and end \texttt{<e>} tokens, respectively, to the tidy
sequence of tokens.

\begin{table}
  \centering
  \begin{tabular}{r | l}
             \texttt{<url>} & replaces a URL beginning with \texttt{http://} or
                              \texttt{https://}\,, \\[0.5em]
             \texttt{<www>} & replaces a URL beginning with the informal
                              \texttt{www.}\,, \\[0.5em]
           \texttt{<email>} & stands for an email address, \\[0.5em]
                            & groups together and replaces three or more\\
    \texttt{<threespecial>} & consecutive non-alphanumeric characters, \\
                            & possibly separated by whitepace. \\[0.25em]
  \end{tabular}
  \vspace{1em}
  \label{tbl:tokens}
  \caption{Special tokens}
\end{table}

The final representation of the email includes only lowercase alphanumeric
words and tokens. Using a list of allowed characters, we aggressively parse the
text, mitigating the threat of the text exhibiting unexpected behaviour.

\subsection{Recurrent Neural Network Classifier}

Our model is a simple RNN, consisting of an encoding layer, two recurrent
layers, and a linear output layer with a Softplus activation. Challenges of
training deep networks, of which RNNs are the deepest, motivate most of the
design decisions presented in this section.

We implement our recurrent layers with the long short-term memory (LSTM)
architecture~\cite{hochreiter1997lstm}. LSTM is a gated recurrent neural layer
architecture that, through its carefully designed self-loops, has the capacity
to learn long range dependencies. We use a variation of the original concept
with weights on the self-loop conditioned on the context~\cite{gers2000lstm}.
Due to its carefully crafted architecture, LSTMs are resistant to the vanishing
gradient problem~\cite{bengio1994difficult}. As is the standard, we use the
tanh nonlinear activation on the cells' output. We describe the choice of the
size of the hidden layer to section below, but we will choose the hidden state
to be 200 variables large.

The output $\bm h_2$ of the last LSTM cell of the second layer is input further
up the model. So that our model outputs a single variable $p_{\widehat y} \in
(0,1)$ as required. Since we are modelling a Bernoulli probability, we use the
simplest linear layer
\begin{equation*}
  \bm h_2 \mapsto \bm w^\intercal \bm h_2 + b = z,
\end{equation*}
consisting of a weight vector $\bm w$ and bias scalar $b$. The final output is
obtained by mapping the linear layer output scalar through the logistic sigmoid
function
\begin{equation*}
  p_{\widehat y} = \sigma(z) := \frac1{1+\exp(-x)} \in (0,1)
\end{equation*}
to obtain the estimated probability of an email being phish.

\subsection{Input Sequence Preprocessing}

If we let every token in the dataset to have its unique embedding vector, not
only would the encoding layer be huge, but our model predictions would not
generalise well to any emails containing unknown words. We hence reduce the
size of the \emph{dictionary} considered by our model, in order to acquire
round values, to the 4\,995 most common words in the training and validation
sets of emails as token sets (\ie, we do not consider repetitions of a word in
a single email in determining the occurrence count).

Every token in the dictionary is assigned a unique index value. So that our
vocabulary reduction is not too harsh, we unite tokens of similar meaning. We
stem the words using the Snowball Stemmer, a more aggressive version of the
popular Porter Stemmer~\cite{snowball}. We then add 5 more tokens
\texttt{<unkalpha>}, \texttt{<unknnum>}, \texttt{<unk>}, \texttt{<cuts>} and
\texttt{<cute>} to the dictionary. The first three abstract out unknown words
to the dictionary, such as those that consist of only alphabetical or numerical
values, or fit none of first two, respectively. We describe the final two
tokens in Section~\ref{sec:cutout} below.

\subsection{Cutout Pruning}
\label{sec:cutout}

Anomalous emails of very long sequence representations cause training
inefficiency, amongst other problems, in evaluating very long range
dependencies. The problem is that such long emails cause unnecessary
\emph{`padding'} of other, shorter sequences, when employing gradient-based
learning in batches, reducing stability and the speed of learning. Most
notably, modern GPU architectures take time proportional to the maximum length
of a sample in the batch to evaluate batched samples, as we do.

We hence compromise our email representation for excessively long emails via a
simple pruning procedure. The idea is to cut out a sequence of size a third of
our threshold of 1000 tokens, and `glue' the beginning and ending of the email
to the \emph{cutout} sequenece. The concept is to keep the beginning, most
middle and ending parts of the email, skipping the uninformative bits of ham or
phish emails. To allow our model to grasp the idea of the anomaly introduced in
close-neighbour word dependencies, we add two tokens, \texttt{<cuts>} and
\texttt{<cute>}, to the dictionary to represent a start or an end of a sequence
caused by the pruning cut. Intuitively, we think of these tokens as
\emph{`glue'}.

Emails represented as sequences of indices of their respective tokens, in the
range of the dictionary size $V = 5000$, are input or \emph{`fed'} to the RNN.
The first, \emph{encoding} layer, encodes each index in sequence with its
corresponding token embedding. The embedding vectors elements are initialised
as random Gaussian $\mathscr{N}(0,0.1^2)$ values and learned as parameters of
the model.

%% file: 5-evaluation.tex
\section{Evaluation}
\label{sec:evaluation}

Before presenting the results of our RNN classifier, we first introduce the
email datasets used in evaluation. Table~\ref{tbl:datasets} presents a summary
of the datasets used. The first dataset, \jnfull, is a combination of all
6\,951 ham emails from the SpamAssassin public corpus~\cite{spamassassin} and
4\,572 phishing emails from the Nazario phishing corpus~\cite{nazario}
collected before August 2007. \jnfull\ is a popular dataset used in related
work to evaluate comparable phishing detection
solutions~\cite{bergholz2008model-based,fette2007learning,verma2012detecting}.

Our second dataset, \enjn, is a combination of the Enron email dataset combined
with phishing emails from the Nazario phishing corpus. The Enron email dataset
is generated by 158 employees of the Enron Corporation, and, to the best of the
authors' knowledge, is the only large public dataset of real-world emails. We
combine a randomly selected subset of 10\,000 emails from the Enron dataset
together with all 9\,962 phishing emails from the Nazario phishing corpus.

\begin{table}
  \centering
  \begin{tabular}{r | r r r c }
    Corpus &
      \multicolumn{1}{c}{Size} &
      \multicolumn{1}{c}{Ham} &
      \multicolumn{1}{c}{Phishing} &
      Source \\ [0.5ex]
    \hline
    \jnfull & 11\,523 &  6\,951 (60\%) & 4\,572 (40\%) & SpamAssassin and
                                                         Nazario \\
      \enjn & 19\,962 & 10\,000 (50\%) & 9\,962 (50\%) & Enron and Nazario
  \end{tabular}
  \vspace{1em}
  \label{tbl:datasets}
  \caption{Decomposition of datasets used in evaluation.}
\end{table}

As is common practice in statistical learning, we split the data samples for
training and evaluation. Separately, we sort the ham and phishing emails by the
\texttt{datetime} stamp extracted from the email \texttt{Received} or
\texttt{Received-Date} field (defined to be the maximum, or latest, timestamp
where multiple \texttt{Received} or \texttt{Received-Date} fields are present).
Consequently, we get two sorted lists, that we separately split into
\emph{training and validation}, and \emph{testing} sets, with a 9-1 ration
twice. The respective 81\% -- 9\% -- 10\% splits respect the received datetime
stamps with the most recent 10\% of the emails forming the training set. The
underlying reasoning is to approximate the real scenario of training the
classifier on present data to predict future data. We then combine the ham and
phishing sets, respecting the splits.

We evaluate our classifier against the most popular metrics in email
classifications, which we introduce shortly. We then compare our language model
to other content-based classifiers.

\subsection{Training}

The encoding itself accounts for $5000 \times 200 = 1\, \mathrm{mil}$
parameters of the model. The challenge of training so many parameters of a
network requires more advanced optimisation algorithms. We employ the following
techniques for optimisation and regularisation of our model.

We initialize the weights of the LSTM cells to random orthogonal matrices with
gain set to $5/3$ for the weights of the cell gate with $\tanh$ activations,
and set the other weights, with sigmoid activations, to orthogonal matrices
with gain~1~\cite{saxe2013exact}. It is the perfect orthogonality of the weight
matrices that motivated our choice for the embedding and LSTM to share the same
unit size of 200.

As suggested by Jozefowicz~\etal~\cite{jozefowicz2015empirical}, we initialize
the bias of the LSTM forget gate to 1, and initialize all other biases to 0
throughout the RNN. We initialise the weights outside of the recurrent layers
by sampling from the Gaussian $\mathscr N(0,0.1^2)$ distribution. The model
contains dropout~\cite{srivastava2014dropout} of $0.2$ on the embedding layer,
a dropout of $0.5$ between all recurrent states on top of each other, with no
dropout in-between successive states of a recurrent layer, as proposed by
Sutskever~\etal~\cite{sutskever2014regularization}. We also add dropout of
$0.5$ at the final output of the recurrent layer.

The model is optimized using the Adam optimizer~\cite{kingma2015adam} against
the binary cross entropy loss function. We train the model with batches of size
200 samples. The training dataset is shuffled at the beginning of every epoch.
To tackle the exploding gradient problem, we clip the gradient norm
$\|\bm{g}\|$~\cite{pascanu2013difficulty} with threshold $1$. Finally, we stop
training early, with continuation of learning~\cite{goodfellow2016book} by
training over the validation set once.

\subsection{Evaluation Metrics}

Given that the datasets used for email classification vary greatly in how even
their distributions are, the obvious \emph{accuracy} measure is of limited
value for comparison to other classifiers. We hence report the standard
measures of \textit{precision}, \textit{recall}, the \textit{$F$-measure},
\emph{false positive} and \emph{false negative rates} in addition to accuracy.

We note that email classification errors vary in importance. As an artifact of
the problem of spam email classification, it is common practice to consider a
false positive error to be more costly than a false negative misclassification.
However, this is under the assumption of aggressive filtering of positives and
harmless false negatives. In the domain of phishing emails, however, false
negatives present significant danger and less aggressive filtering methods such
as alerts and link-disabling are common.

We train the classifier over 4 epochs on the training dataset and 1 more epoch
over the validation dataset. Because the model is expensive to train, in time
and computational power, the results provided are of the single trained
instance. We evaluate the model on the test set, which had been unseen during
training, and is chronologically separated from training a validation set. This
is due to the fact that we split each dataset into training, validation and
testing sets in chronological order.

Our classifier is most directly comparable to other text-based features, or
sub-classifiers that analyse the text of the classified email only. We compare
our work with the $\nettext$ sub-classifier of the PhishNet-NLP email
classifier by Verma~\etal~\cite{verma2012detecting}, and the state-of-the-art
\emph{Dynamic Markov chain} (DMC) model proposed by
Bergholz~\etal~\cite{bergholz2008model-based}. We summarise the results in
Table~\ref{tbl:results}.

\begin{table}[h]
  \resizebox{\textwidth}{!}{%
    \begin{tabular}{r | c | rrrrrrr }
      & corpus & accuracy &  fp-rate &  fn-rate & precision &   recall
        & $F$-measure \\ [0.5ex]
    \hline
    {\scriptsize\nettext}
      &  \pnjn  & 78.54 \% & 14.90 \% & 22.90 \% &  95.93 \% & 77.10 \%
        & 85.49 \% \\
    \dmctext
      & \jnfull & 99.56 \% &  0.00 \% &  4.02 \% & 100.00 \% & 95.98 \%
        & 97.95 \% \\
    our RNN
      & \jnfull & 98.91 \% &  1.26 \% &  1.47 \% &  98.74 \% & 98.53 \%
        & 98.63 \% \\
    \hline
    our RNN
      &  \enjn  & 96.74 \% &  2.50 \% &  4.02 \% &  97.45 \% & 95.98 \%
        & 96.71 \%
    \end{tabular}%
  }
  \vspace{1em}
  \caption{%
    Summary of our results in comparison to related work in popular metrics.%
  }
  \label{tbl:results}
\end{table}

Our test dataset is well-separated from the training set. We could argue that
the classification problem we evaluated our classifier against is
unrealistically hard. Intuitively, messages arriving to a specific inbox would
exhibit more pronounced patterns, and would thus be easier to classify
correctly.

Verma~\etal~\cite{verma2012detecting} propose that $\nettext$ offers a
classification value very independent from the other features, as only the text
of the email is considered. For the same reason, our classifier should not copy
the labels of other features present in classification, but rather provide an
independent view on the classification at hand.

The RNN classifier clearly outperforms the $\nettext$ classifier, and has
comparable results to the state-of-the-art $\dmctext$ feature. We note that
perfect classification is not possible in our setting, as two emails with the
same token sequence will necessarily be labelled equally. Since both, ham and
phishing email corpus contain empty emails with attachments, which have been
removed, the emails are identical to our classifier. This proves inseparability
of the emails with the word-sequence representation.

%% file: 6-conclusion.tex
\section{Conclusion}
\label{sec:conclusion}

In this paper, we propose a novel automated system aiming to mitigate the
threat of phishing emails with the use of RNNs. Our results suggest that the
flexibility of RNNs gives our system an edge over the expert feature selection
procedure, which is vastly employed in Machine-Learning-based attempts at
phishing mitigation.

We focused on the overlooked content source of email information and
demonstrated its utility when considered in phishing threat mitigation. The
nature of RNN and its training procedure make it suitable for the case of
online learning deployment. Our classifier could theoretically change over time
to capture new trends continuously and keep up accurate and precise
classification throughout. Our results have demonstrated a wealth of potential
in non-trivial feature identification for classifying emails, since oru
system's performance surpasses the state-of-the-art systems which are based on
features designed by human intuition.

Finally, it is worth noting that the general criticism of supervised learning
extends to our case. Little information is provided by the RNN classifier on
the nature of emails at successful classification. The proposed solution
generalises easily to the case of inclusion of basic spam emails, and is a
prospect for further automated success.